\documentclass{article}

\usepackage{calculator}
\usepackage{url}
\usepackage[utf8]{inputenc}
\usepackage{authblk}
\usepackage{amsmath}
\usepackage{xcolor}
\usepackage{graphicx}
\newcommand{\ignore}[1]{}

\newcommand{\rev}[1]{{\color{black}{#1}}}
\usepackage[margin=1in]{geometry}

\begin{titlepage}

\title{Hybrid Forecasting of Geopolitical Events\footnote{In memoriam of Michael D. Ward. This paper would not be possible without his contributions.}}

\date{February 18, 2023}

\author[1,2]{Daniel M. Benjamin}
\author[1]{\; Fred Morstatter}
\author[3]{\; Ali E. Abbas} 
\author[1,4]{\; Andres Abeliuk}
\author[5]{\; Pavel Atanasov} 
\author[6]{\; Stephen Bennett}
\author[7]{\; Andreas Beger}
\author[1]{\; Saurabh Birari}
\author[8]{\; David V. Budescu}
\author[9]{\; Michele Catasta\footnote{Currently at Replit}}
\author[1]{\; Emilio Ferrara}
\author[3]{\; Lucas Haravitch}
\author[8]{\; Mark Himmelstein}
\author[1]{\; KSM Tozammel Hossain\footnote{Currently at University of Missouri, Columbia, MO}}
\author[1]{\; Yuzhong Huang}
\author[1]{\; Woojeong Jin}
\author[10]{\; Regina Joseph}
\author[9]{\; Jure Leskovec}
\author[1]{\; Akira Matsui\footnote{Currently at Yokohama National University, Kanagawa, Japan}} 
\author[1]{\; Mehrnoosh Mirtaheri}
\author[1]{\; Xiang Ren}
\author[1]{\; Gleb Satyukov}
\author[11]{\; Rajiv Sethi}
\author[1]{\; Amandeep Singh}
\author[9]{\; Rok Sosic}
\author[6]{\; Mark Steyvers}
\author[1]{\; Pedro A Szekely} 
\author[7]{\; Michael D. Ward}
\author[1]{\; Aram Galstyan\footnote{Corresponding author: { galstyan@isi.edu} }}

\affil[1]{USC Information Sciences Institute, Marina del Rey, CA}
\affil[2]{Nova Southeastern University, Fort Lauderdale, FL}
\affil[3]{University of Southern California, Los Angeles, CA}
\affil[4]{University of Chile, Santiago, Chile; National Center for Artificial Intelligence (CENIA), Chile}
\affil[5]{Pytho, LLC, New York, NY}
\affil[6]{University of California Irvine, Irvine, CA}
\affil[7]{Predictive Heuristics, Seattle, WA, USA}
\affil[8]{Fordham University, Bronx, NY}
\affil[9]{Stanford University, Stanford, CA}
\affil[10]{Sibylink, New York, NY}
\affil[11]{Barnard College, Columbia University, New York, NY}

\end{titlepage}

\begin{document}

\maketitle


\begin{abstract}
Sound decision-making relies on accurate prediction for tangible outcomes ranging from military conflict to disease outbreaks. \rev{To improve crowdsourced forecasting accuracy,} we \rev{developed} SAGE, a {\em hybrid forecasting system} that combines human and machine generated forecasts. The system provides a platform where users can interact with machine models and thus anchor their judgments on an objective benchmark. The system also aggregates human and machine forecasts weighting both for propinquity and based on assessed skill while adjusting for overconfidence. We present results from the Hybrid Forecasting Competition (HFC) -- larger than comparable forecasting tournaments -- including 1085 users forecasting 398 real-world forecasting problems over eight months. Our main result is that the hybrid system generated more accurate forecasts compared to a human-only baseline which had no machine generated predictions. We found that skilled forecasters who had access to machine-generated forecasts outperformed those who only viewed historical data. We also demonstrated the inclusion of machine-generated forecasts in our aggregation algorithms improved performance, both in terms of accuracy and scalability. This suggests that hybrid forecasting systems, which potentially require fewer human resources, can be a viable approach for maintaining a competitive level of accuracy over a larger number of forecasting questions. 
\end{abstract}

\section{Introduction}
From military conflicts to disease outbreak to economic disruption, accurate prediction is vital for sound intelligence-based decision-making. However, the problem of making accurate predictions for geopolitical events is notoriously difficult due to too much or too little data, rare event occurrences, or large levels of uncertainty. Prediction methods range from expert and/or group judgment to individual and ensembled statistical models~\cite{zellner_survey_nodate}. It is often challenging to identify a consistent, superior prediction method~\cite{meehl_clinical_1954, zellner_survey_nodate}. Stakeholders confidently misidentify the benefits of competing methods, such as trusting human clinical judgment over statistical or algorithmic judgment~\cite{dawes_clinical_1989}. Two common forecasting methods, crowdsourcing and machine learning, have complementary strengths and competing weaknesses. Here we present a hybrid forecasting model - a system that aims to exploit the proficiencies of each while circumventing their deficiencies. 

Recent forecasting tournaments such as IARPA's Aggregative Contingent Estimation (ACE)~\cite{noauthor_iarpa_2011} have led to advances in crowdsourcing methods, statistical aggregation, and ultimately improvements in accuracy~\cite{atanasov_distilling_2017,mellers2014psychological}. Crowdsourced aggregation pools a breadth of knowledge while canceling independent errors~\cite{budescu_confidence_2006} and is most successful when individual performance can be tracked over time. However, social influences can harm opinion pools, and individual (rewards) vs. group incentives can be difficult to balance. Individuals must choose to share their private information and trust others. 

Advances in statistical and machine learning methods lead to accuracy gains due to their ability to handle troves of data with heterogeneous input and identify complex relationships~\cite{ghoddusi_machine_2019}. Data-driven, algorithmic forecasting can be used to predict various political outcomes, such as terrorism, conflict, insurgency, and similar~\cite{enders_patterns_2002, pilster_predicting_2014, schrodt_automated_2010, schutte_regions_2017}. 
However, machine learning requires large amounts of data to be available and accessible. If data is not in a standard format, there can be large costs to pre-processing data. 

Statistical models perform well under the right circumstances~\cite{leigh_competing_2006}, and human crowds succeed when deftly combined~\cite{atanasov_distilling_2017}. Factors like amount, availability, and structure of data determine how these methods perform~\cite{seifert_3_2013}. Machine-based forecasting methods typically perform well on problems for which there is sufficient historical data, but are ill-suited to forecast rare or idiosyncratic events for which such data may not exist, or when the underlying context has changed in ways not reflected by the historical data. \rev{Machine predictions handle data in a consistent, structured manner and avoid computational errors, like violating probabilitiy axioms~\cite{dellermann_design_2019}.}

Human analysts, on the other hand, can often accurately forecast outcomes without exclusively depending on availability of historical data, by leveraging their domain knowledge and prior experience. \rev{Further, human expertise and domain knowledge can be valuable as inputs into machine models. These benefits are most efficient when data is sparse and/or unstructured~\cite{dellermann_design_2019}.} However, even the best analysts may not match machine performance where solid historical data is available and can be cognitively overwhelmed when addressing a large number of problems within time constraints, thereby limiting the scalability of a forecasting system that relies solely on human judgment. Unfortunately, there are few direct comparisons between models and crowds in similar settings. 

Here we describe our Synergistic Anticipation of Geopolitical Events (SAGE) system, which was developed under IARPA's Hybrid Forecasting Competition (HFC) program~\cite{noauthor_iarpa_2017}. \rev{The system is designed to make verifiable probabilistic predictions of outcomes from a broad set of domains, such as politics and international relations (ie the quantity of battle deaths or piracy in a region or attributable to a specified actor), health and disease (ie flu or dengue fever case counts), economics and finance (ie exchange rates or oil prices), and science and nature (ie the number of earthquakes or cybersecurity breaches) (see section \ref{IFPs} for an overview of the types of questions). A human-computer system can achieve ``hybrid intelligence'' when applied in a setting with a high degree of digitization and human expertise~\cite{rafner_revisiting_2021}.} SAGE is a hybrid forecasting platform that allows human forecasters to combine model-based forecasts with their own judgment. The SAGE system provides forecasters automated statistical predictions and freedom to choose if and how much weight to assign to model predictions when submitting their personal forecasts \rev{since formal models can increase the skill of human judges~\cite{rafner_revisiting_2021}.} 

Our system is designed to \rev{test the conceptual hypothesis that machine model forecasts embedded in a crowdsourced forecasting platform can improve the accuracy and efficiency of established crowdsourced forecasting methods. We embed machine models in a system designed to} balance a) the diversity required to achieve the ``wisdom of the crowd'' by not restricting users' responses with b) anchoring forecasters to an impartial benchmark to minimize noise and outliers. \rev{ This paper tests \emph{how} machine models lead to improvements. We experimentally test various informational conditions to determine which type of information -- historical data, model output, or interactivity with the models -- leads to optimal accuracy and user engagement (see section \ref{methods} for details). We also test if machine models improve system efficiency. We hypothesize that machine models can help increase the number of questions SAGE can answer without decreasing accuracy. We test methods of allocating a fixed number of human forecasters to questions where they are most needed. 

In what follows, we test if our hybrid system can improve accuracy, engagement, or scalability compared to established crowdsourcing methods.} \rev{First, we discuss relevant literature related to hybrid intelligence and scalability. Then, we} describe the main components of SAGE \rev{followed by a description of HFC guidelines. Finally, we}present our experimental results from a 8-months long Randomized Controlled Trial (RCT) conducted under the HFC program.

\section{Related Works}

\rev{
The main motivation behind developing a ``hybrid'' forecasting system is to harness the strengths of crowdsourced and statistical forecasts by combining them with machine learning models as input for both human forecasters and aggregation methods. This type of \emph{hybrid intelligence} occurs when human and machine components each contribute to a solution that outperforms and/or is more efficient than either source on its own~\cite{dellermann_design_2019, kamar_hybrid_2016}. Machine models, which excel at identifying patterns from data and leveraging them for making predictions, can help human judges overcome certain errors and inconsistencies. Human experts, which do not require structured input data, are capable of ad hoc feature selection, often quicker than variables can be formalized when data sources are yet unavailable.  

While there is a growing field discussing the current state of hybrid intelligence, there is limited work exploring how such systems work and in what settings they excel. To date, most work explicitly discussing hybrid intelligence is theoretical (e.g. ~\cite{dellermann_developing_2017, rafner_revisiting_2021, rafner_mapping_2022}). Developing an efficient and effective hybrid system to solve complex, dynamic tasks requires a carefully designed and tested machine component, a skilled human component, and principled, dynamic methods for combining them. In the current study, we address the challenges of balancing effectiveness with flexibility. Artificial intelligence exceeds when tasks are well-defined (e.g. ~\cite{dellermann_design_2019}). Machine models can underperform when tasks are loosely defined, data is sparse, or environments are complex and/or changing. A forecasting tournament provides an opportunity to collect data in a structured, yet chaotic, environment. On-one-hand, the general question and response format is consistent and practiced users provide consistent response data. On-the-other-hand, it is a difficult setting to generalize because new question types, sources, and datasets could be introduced after system development. 

One key limitation of the previous work on crowdsourced forecasting and hybrid intelligence is that use cases are limited and often applied to a business environment (e.g. ~\cite{dellermann_design_2019, dellermann_developing_2017, rafner_deskilling_2022}). The current study is designed to provide data-driven support for the effectiveness of a hybrid system in the geopolitical forecasting domain. While there are established methods showing how crowdsourced forecasting succeeds, there are not established methods for a hybrid forecasting system~\cite{mellers2014psychological}. Previous crowdsourced methods rely on adjustments, such as statistical recalibration, to adjust for measurable biases in human forecasters like overconfidence~\cite{atanasov_distilling_2017}. It remains an open question whether the same or new cognitive biases emerge when human users interact with machine model output. Research suggests that presenting time-series data as a forecasting aid improves individual forecasts by reducing random error~\cite{de_baets_forecasting_2018}. When there are detectable trends in a time series, forecasts made while viewing graphical data are more accurate than from viewing tabular data~\cite{harvey_graphs_1996} . In this setting, a hybrid system must account for potential, yet unmeasured, biases to effectively combine machine models and crowd predictions. 

The success of machine models is driven by complexity and volatilitiy. When predicting on real data, machine learning models face a tradeoff between the complexity of the data and the number of model parameters required to predict accurately~\cite{parmezan_evaluation_2019}. Hence, tuning and re-turning becomes cumbersome when properties of the event or dataset change. The problem becomes more challenging when predicting several periods into the future and requires methods that produce multiple outcomes~\cite{ben_taieb_review_2012}. Known (simple) statistical methods often outperform more sophisticated methods (e.g. based on deep neural networks) because real-world time-series are often non-stationary. Changes from training to testing often impede how well sophisticated models generalize~\cite{makridakis_statistical_2018}. It is yet not known whether human judgment can help identify the shifts in time-series over time that make statistical and machine model predictions miss the mark. Further, while the main focus of the machine models considered here is on quantitative, time-series data, there is also an emerging line of work which intends to use unstructured textual data for making predictions. For example, ~\cite{Hossain2022} aims to extract possible precursors of certain events from documents, while ~\cite{jin-etal-2021-forecastqa} formulates forecasting as a Question Answering (QA) problem on an appropriately selected textual dataset. 

A key aspect in achieving an efficient and effective hybrid system is how to allocate both the human and machine resources. Intelligent task allocation can bring out the best in both sources (e.g. in classification~\cite{beck_integrating_2018}; in consensus~\cite{kamar_hybrid_2016}. In a review of 208 articles over 50 years, task allocation is identified as one of the key issues to making hybrid system work~\cite{janssen_history_2019}. It is challenging to allocate tasks When it is not knowable in advance at which tasks machines and humans will outperform each other. As the task becomes more difficult and the system becomes more complex, task allocation becomes more difficult~\cite{trouille_citizen_2019}. Introducing machine elements into crowd systems comes with trade-offs. Misallocation can diminish engagement and shift attention away from desired tasks. The ``wisdom of the crowd'' effect relies on sufficient expertise and diversity of knowledge. In this setting, the introduction of statistical models could diminish diversity if human participants are too trusting in the models and do not feel empowered or motivated to add their private information into the system. Intelligent task allocation must balance finding the best individual sources for a given task with maintaining a diverse pool of knowledge.  
}

\section{Methods} \label{methods}
\subsection{SAGE System}
The Synergistic Anticipation of Geopolitical Events (SAGE) system was developed to combine automated statistical forecasts with a pool of human knowledge by allowing users access to machine model output and by algorithmically combining human and machine forecasts~\cite{ijcai2019-955}. The SAGE platform allowed users to interact with machine models to anchor their judgments on an objective benchmark. Simultaneously, users had the freedom to choose if and how they combined model forecasts with their own judgment striving for the diversity of knowledge needed for the ``wisdom of the crowd'' effect~\cite{abeliuk_quantifying_2020}. To proactively mitigate skepticism with and over-reliance on the models, we trained users in how to evaluate and consolidate information from multiple sources. 
\begin{figure}[!h]
    \centering
    \includegraphics[width=\textwidth]{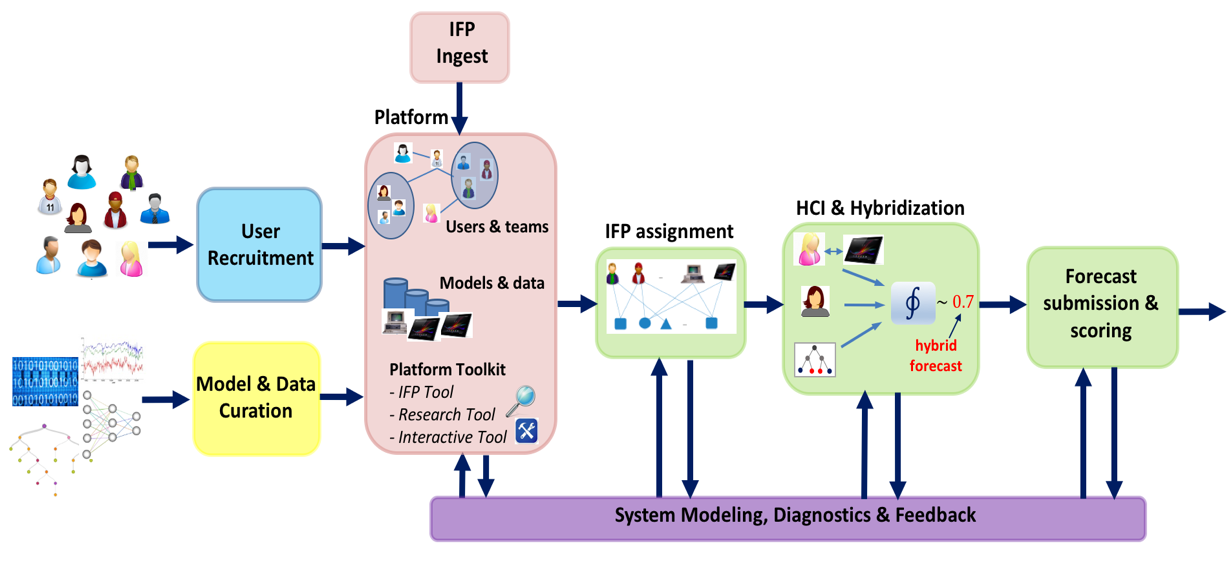}
    \caption{Schematic of SAGE system organized into five topic areas. Platform engineering is in pink, recruitment and retention is in blue, machine-based forecasting is in yellow, human-machine interaction is in green, and diagnostics and feedback is in purple.}
    \label{fig:schematic}
\end{figure}

The SAGE system was developed by integrating five areas of engineering and design (see Figure \ref{fig:schematic}). After logging both machine and human forecasts, our aggregation algorithms computed aggregate forecasts in real time by dynamically combining human and machine forecasts. Over time, our system determined optimal weights for each source based on assessed skill, adjusting for overconfidence, and for propinquity to question resolution. We also developed a number of machine models allowing the system to choose reliable models for various question types. Finally, our system adaptively filtered questions balancing users' preferences and abilities with the systems' needs. Our recommender system filtered questions to the top that individuals were more likely to answer and/or were unpopular, while hiding questions that were overly popular.

\subsection{Forecasting Platform}
\begin{figure}[!h]
    \centering
    \includegraphics[width=\textwidth]{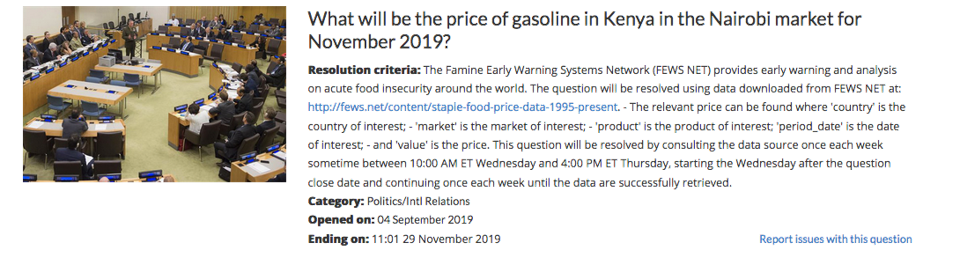}
    \caption{Screen capture of an IFP with resolution criteria.}
    \label{fig:ifp}
\end{figure}

The SAGE system included search and filtering functionality to help users find forecasting questions (IFPs) about which they felt knowledgeable. At any point in time, there were dozens of IFPs available. Users had to complete at least 5 forecasts per week; An example of an IFP is shown in Fig.~\ref{fig:ifp}. After choosing which question to answer, an IFP page included the following from top to bottom: question text, resolution criteria (including the source used to resolve a given question, value of interest and/or criteria for an occurrence, and timing), automated information including data graph, statistical forecast, and interactive features (depending on their experimental condition), forecast sliders that forced responses to add to 100\%, a textbox to justify the forecast, and a comments thread to view and respond to fellow users' justifications. Additional features included a leaderboard, consensus charts, a research tool, a profile page including their personal accomplishments, and training and tournament information.

\subsection{Data Pipeline and Model Development}\label{sec:methods-models}

The SAGE machine model pipeline can be broken down into two parts based on the kinds of questions that were covered. Approximately 45\% of IFPs (AKA data-driven IFPs) were clearly associated with a univariate time series, like OECD interest rates for a country\cite{OECD_nodate}. These questions were covered by automated data-acquisition and univariate time-series forecasting systems. The remaining questions did not have clearly associated data. Some, about election results and country leader resignations, were covered by tailored models that could leverage more complicated, non-time-series data. Others were covered by tools that leveraged resolved answers to other, similar, previous questions, or extracted relevant information from the ICEWS event data~\cite{boschee_icews_2022}. The non-time series models either only covered a very small set of questions, or did not perform well in terms of accuracy, so the rest of this section will focus on the time series forecasting system.

The time series forecasting system consisted of a data platform that maintained a continuously updated database of relevant time series data sets and could map them to questions as appropriate, and a forecasting platform that would then parse a question and apply a univariate time series model to derive probabilities for the question answers. Our system was developed to automate data extraction based on reading the question text and finding the applicable data. Many data sources were known in advance and several were not. 

A significant challenge was to identify the time series models to use for generating the forecasts that would be shown to users and sent to the aggregation models. \rev{Four core models were displayed to users in the question charts: the auto ARIMA model, a similar automated exponential smoothing model (ETS)~\cite{hyndman2018forecasting, hyndman_automatic_2008}, a simple random walk model, and the M4-Metalearning model~\cite{montero2020fforma}. The DCT Ensemble model, drawing on forecasts from auto ARIMA, M4-Metalearning, or a AR(1) neural net model based on an analysis of the input series discrete cosine transformation was used to provision forecasts for aggregation. Model performance suffered from a "cold-start problem" as the number of IFP resolutions were limited over the first several months of the competition. Further, HFC guidelines required our system to make predictions for new datasets and sources on the fly, often with only a couple hours notice before users could access the IFPs. Therefore, simple time-series models tended to outperform more complex, topic-specific models -- a result supported by the general success of conservative forecasting approaches~\cite{armstrong_golden_2015, makridakis_statistical_2018}. } Initially all forecasts were based on the Auto ARIMA model~\cite{hyndman2018forecasting}, but later this was supplanted by an ensemble (labelled ``PHE2'' below) of Auto ARIMA and an exponential smoothing state space model~\cite{hyndman2018forecasting}, which emerged from an overall pool of 28 candidate models. \rev{We do not report results from poor performing models.} Choosing adequate models was hard because inter-question performance a is very noisy (variable), yet only relatively small numbers of resolved questions were available for testing \rev{and several models did not have adequate information to specify them consistently.} Only later did enough resolved questions accumulate for model-to-model performance to stabilize.

\subsection{Experimental Conditions}

We conducted a controlled experiment to better understand the benefits of exposing forecasters to different hybridization components. We randomly assigned our 547 participants to one of three experimental conditions which we labeled B, C, D to reflect the increasing level of complexity of and interactivity with the hybridization model. 
\begin{enumerate}
    \item \textbf{Condition B:} This condition exposed users to historical data about the target item. Data included relevant news articles from the research tool, and historical figures that pertains to the question. Historical charts were available for 177 of the 398 items.
    \item \textbf{Condition C:} This condition supplemented the data charts from Condition B with machine model predictions, when available. More specifically, we exposed the forecasters to predictions from the ARIMA model, which has been determined to be a good general model. ARIMA model predictions were available for 177 of the 398 items. 
    \item \textbf{Condition D:} This is a variation on condition C that allows the forecasters to tweak the parameters of the visualization, including the type of model and range of data used for model training. We also provided a simple method that allowed the judges to adjust the model’s forecast by selecting the mean and variance of the target value and directly translating that into a forecast\footnote{Behavioral decision making researchers have repeatedly documented a pattern of “Algorithm Aversion (AA for short)”  (e.g.,~\cite{burton_systematic_2020, dietvorst_b_j_algorithm_2015})) - the tendency of humans to prefer and value advice and information from human sources over machine counterparts, even when the information provided by humans and algorithms is identical (e.g.,~\cite{dietvorst_b_j_algorithm_2015, onkal_relative_2009}).  In general, judges tend to be less tolerant of errors made by algorithms, compared to humans (e.g.,~\cite{dietvorst_b_j_algorithm_2015, prahl_understanding_2017}).  One way to reduce AA is to allow people to have more control over the algorithm by tweaking it, or some of its predictions~\cite{dietvorst_overcoming_2018}. Condition D was implemented to test this expectation. }.
\end{enumerate}
\begin{figure}[!h]
    \centering
    \includegraphics[width=\textwidth]{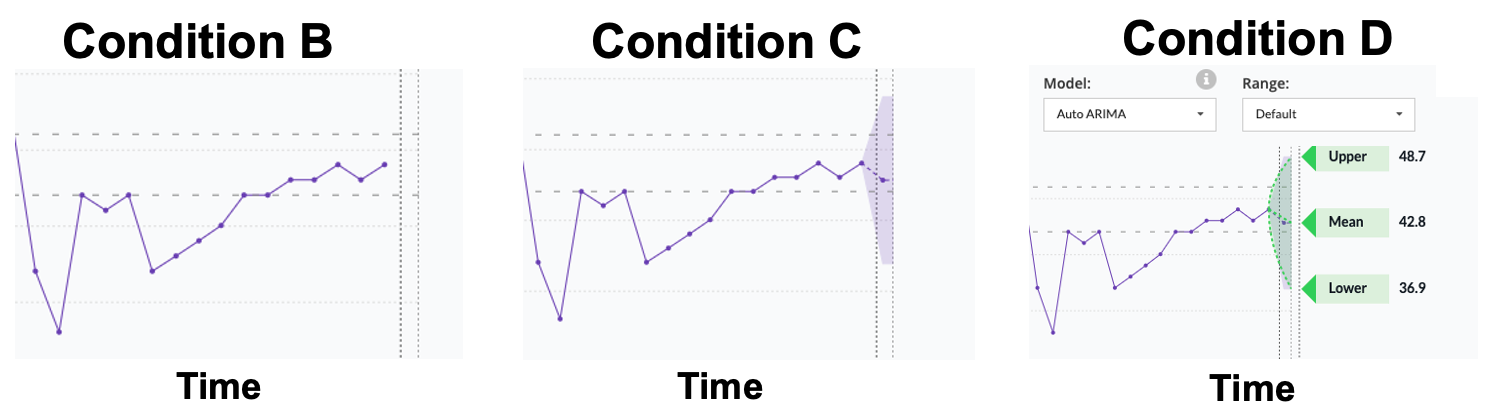}
    \caption{Schematic illustration of information presented to participants in each experimental condition.}
    \label{fig:conditions}
\end{figure}
Figure \ref{fig:conditions} presents examples of screenshots from Conditions B, C and D. \rev{Control (see below) and} Condition B quantified the ability of human forecasters to predict the various items and provide natural baselines to compare forecasters in Conditions C and D that had access to machine models. The benefit of this access is measured by the improvement in accuracy, compared to the controls.  
In addition to the three {\em treatment} conditions above, there was a control condition that was run separately by the HFC Test and Evaluation Team. This control condition used a different platform with a different sample of 538 respondents selected from the same pool. The control condition did not offer any historical charts nor machine predictions to the participants. The main objective of the program was shown that the {\em hybridized} conditions could generate more accurate aggregate forecasts than the control.

\subsection{Forecast Aggregation} \label{agg methods}
By combining human and machine model forecasts, we aimed to leverage the collective intelligence of human and model judgments. The advantage of human judgment was its flexibility and ability to reason with qualitative and mixed-source data. Humans can forecast when data is sparse or difficult to interpret and can seek out information that only indirectly relates to the question at hand. On the other hand, the advantages of models included expeditious forecasting which improved scalability. Statistical models dutifully forecasted on any number of questions and their accuracy tends to improve as more data became available. 

The key challenge for aggregation was that many factors related to human and model judgment were not known \textit{a priori}. For each particular forecasting problem, the total number of human forecasts was not knowable in advance. On any particular day the forecasting problem was available, a handful of human forecasts might be produced but in some extreme cases, no human judgment might be available for the entire duration of the forecasting problem. In addition, at the start of the forecasting project, it was not known what the relative accuracy is of the model and human judgments for certain types of forecasting problems. Every introduction of a new type of forecasting problem injected new uncertainty about the relative capabilities of human and model forecasting accuracy. This cold-start problem made it challenging to apply machine-learning approaches that can learn optimal combinations of human and model judgment as large quantities of human judgments were initially not available and yet accurate forecasts needed to be produced from the start of the project. Therefore, the goal for aggregation was to develop a robust framework for integrating human and model judgment with the potential to scale to large numbers of forecasting questions. 

To combine the human forecasts, we employed a combination of tested methods and new strategies to maximize performance. The aggregation of human-only forecasts accounted for three factors: recency, individual skill, and miscalibration. First, our algorithms diminished each forecasts’ value over time as new information accumulates. To account for this recency effect, we kept only the most recent 40\% of forecasts for a question at any given time, and further applied exponential decay to down-weight older forecasts included in the aggregation. Second, we placed higher weights on forecasters forecasters with better accuracy track records, those who updated their forecasts in frequent, small increments~\cite{atanasov_small_2020}, and those who wrote longer text rationales, with more sources and quantitative information. Finally, we recalibrated forecasts to correct for the general tendency toward overconfidence by individual forecasters, and underconfidence of aggregated crowd judgments, especially when aggregated using the mean. This was done by making forecasts by individual forecasts less extreme, but aggregate-level forecasts more extreme (closer to 0\% or 100\%). The overall effect was to make final aggregate estimates slightly more extreme than the equivalent estimates with no recalibration. The best-performing slot used a variant of this aggregation model which made a) forecaster weights more unequal over time, and b) extremization parameters larger over time, making season-end aggregated forecasts more extreme than those at season-start.

Human forecasts were then combined with machine model estimates. Each model forecast was also given weights based on the historical performance of the model that generated it. Our initial strategy for human-machine aggregation was to assess machine weights relative to those of crowd estimates (e.g., a model estimate may be weighted 1/4 as much as the crowd). The more advanced alternative that was used in most slots in the last season (including the best-performing slot), placed weights on model forecasts equal to those of several average-skill individual human forecasters. Initial human-machine weights were set based on backcasting analyses, and were allowed to vary over time based on relative within-season performance in some slots. As a consequence, the aggregate forecast was heavily weighted toward when few human forecasters had placed estimates on the question, when the human forecasts were out of date, or when most active human forecasters on the question were considered low-skill. \rev{We also tested more sophisticated ensemble aggregation methods which used multiple machine models as inputs as well as machine learning-based aggregation which included additional inputs such as the statistical traits of the community forecast, linguistic features of the IFPs and forecast justifications, etc. In a separate paper, we present a neural machine translation aggregation method which assigns anchor attention weights to forecast-user-datetime combinations~\cite{huang_anchor_2022}. We report results from the best performing aggregation method that was run in real-time during the forecast season below, which is based on the method described above. }

\subsection{Training} 
We sought to understand whether training could improve predictive accuracy under conditions in which forecasters had to balance trust in a model with their own judgment. Our HABIT training method combined probabilistic reasoning with hybridization concepts using a character-based narrative device rendered in a cartoon format. \rev{Our training was designed to extend previous vignette-based methods, which focused on core tenets of probabilistic estimation~\cite{mellers2014psychological}, to teach about the machine models involved in the hybridized ensembles and aggregations and how to integrate model forecasts with one's personal knowledge.} We hypothesized that the cognitive burden of whether to integrate or reject machine model data could be mitigated by briefly explaining how each model works and how to balance too little and too much trust in models. We tested both whether or not mandating training improved accuracy and whether the presentation format, whether animated or static, led to gains.

\subsection{Matching Participants with Forecasting Problems}
\label{sec:matching}
The SAGE system aimed to optimize two seemingly conflicting objectives: 1) allow users choice of questions based on their expertise and interests with 2) timely coverage of all questions with limited human forecasters. We developed an IFP Recommender System which presented a personalized ranking of the IFPs on the Question page, based on the specific characteristics of a forecaster. \rev{We develop a recommender system based on the wide and deep learning model~\cite{cheng_wide_2016}. This model identifies preferences using known IFP features, and generalizes to other IFPs via IFP embeddings.} As features, we took into account the performance of the given forecaster on similar past IFPs (based on an IFP Semantic similarity model we developed using BERT~\cite{devlin_bert_2019}) and the user activity on the other IFPs (based on a collaborative filtering scheme). \rev{When designing the SAGE recommender system, BERT proved to be the more accurate, most efficient model because it captured the subtleties in the differences between IFP texts. To gauge the similarity among IFP texts, we used cosine similarity, a common distance metric used in embedding spaces.} We balanced individual with system performance by also capping popular IFPs where consensus was already reached, freeing up human resources to forecast on other IFPs.

\subsection{Data Availability}
SAGE platform data including machine model output and experimental user forecasts, activity, and scores can be found on the Harvard Dataverse ~\cite{morstatter_rct-b_2021}. Control user forecasts, question metadata, and resolutions can be found on a separate Harvard Dataverse page ~\cite{hilliard_hfc_2020}.

\section{HFC Background and Rules}
IARPA's Hybrid Forecasting Competition (HFC)~\cite{noauthor_iarpa_2017} was a multi-year research program developed to test if and how machine-models could improve upon previous crowd-sourced geopolitical forecasting tournaments such as ACE~\cite{noauthor_iarpa_2011}. As stated in the program announcement, ``{\em the goal of HFC was to integrate the strengths of human cognitive and reasoning abilities with those of machine-driven systems to produce maximally accurate forecasts of geopolitical and economic events}''~\cite{noauthor_iarpa_2017}. The evaluation of hybrid forecasting system was conducted via Randomized Controlled Trials (RCT-s). There were two RCT-s during the lifetime of the HFC program. Here we focus our analysis on the second evaluation, referred to as RCT-B, which took place from April to November 2019.

Like all forecasting tournaments, competitors must abide by rules provided by the sponsor and test and evaluation teams. Some rules governed how forecasting questions were developed, to which datasets they were linked, and how and when they would be resolved and scored. Some rules governed the activities of the human users including how they were recruited and assigned to competitor teams. Other rules governed the development and upkeep of machine-model components as well as how the machine models could be combined with human forecasts and when responses must be submitted.

\subsection{Individual Forecasting Problems} \label{IFPs}

During RCT-B, forecasts were conducted on 398 questions, broadly referred to as Individual Forecasting Problem (IFP). The questions covered a broad set of domains, such as politics and international relation, science, health and disease, microeconomics and finance (see ~\cite{hfc-dataverse} for a detailed description of different IFP types).
New IFPs were published on the same day each week. Each IFP was associated with $C$ mutually exclusive and exhaustive outcome events, where 2 $ \le $ C $ \le $ 5. Participants submitted their forecasts for a given IFP by entering a probability for each outcome, where the probabilities across all $C$ outcomes were required to total 100\%. All IFPs had a start date and an end date during which participants could make forecasts for that questions as often as they liked (see Figure \ref{fig:ifp} for a screen capture). IFPs ranged from 2 weeks, to the full 8-month season in duration. IFPs had a mean duration of 87.07 days (SD = 55.85). 205 IFPs had only two response options, the remaining 193 had more than 2 possible responses. Of these 193, 154 were ordinal, in that there was a meaningful ordering to the C events, while the remaining 39 were nominal.

\subsection{Participants}
Human participants were recruited via Amazon Mechanical Turk. CloudResearch filtered the participants to ensure a high level of engagement both prior to the start of the forecasting season by only including users with longitudinal study experience, and mid-season by removing users with low quality responses by assessing the content of their justifications~\cite{moss_how_2022}. The sample consisted of 547 participants, 229 women (42\%), with a mean age of 36.68 (SD = 10.88). A forecasting session consisted of weekly Human Intelligence Tasks (HITs), where each forecaster was required to make at least five forecasts. If possible, three of these five were required to be updates of previous forecasts. For each completed HIT, Participants were paid \$20 per HIT. Participants were permitted to make additional forecasts beyond these five but were not paid for these additional forecasts. Participants were also eligible for accuracy awards if they participated enough. They could earn a portion of a fixed prize pool at the midway and final points. The pool was divided among three prize tiers of \$200, \$100, or \$50 for observed accuracy - as measured with mean daily Brier scores.

\subsection{Machine Models}
A key component of the HFC was the requirement for systems to produce model-based forecasts. In previous comparisons of human and model forecasts, the latter were generated in a traditional fashion by analysts (e.g.~\cite{tetlock2017expert}). In contrast, model-based forecasts for the HFC competition had to be generated by an automated system with restrictions on manual interventions into the process. Fixing system issues, i.e. bugs and similar errors, was allowed, but manual model development like deciding what data and model(s) to use for a question, tuning model parameters, etc. was not permitted. Some data sources were introduced mid-season. Sometimes notice about new data sources came only a couple hours prior to the associated IFPs getting published for human responses. Thus, performer teams were required to quickly produce model forecasts to aid users, and there was insufficient time to build specialized models tuned to specific datasets.

\subsection{Response Submissions}
Each team was allotted 40 official and up to an additional 60 experimental slots for submitting forecasts on each IFP. These slots allowed teams to test multiple theoretical ideas as well as fine-tuning the application of those ideas, such as including inputs or tuning key parameters in systematic ways. A total of ten official slots were locked prohibiting changes by performers, and 30 were unlocked allowing for changing to key model parameters. The experimental slots encouraged testing more novel, higher risk ideas. Performer teams were required to submit one forecast per IFP per submission slot from the day an IFP was originally published to the day it resolved, either on its stated resolution date or due to an event occurrence. For each slot, performer teams had to develop aggregation algorithms that combined the various machine and human inputs and IFP metadata for a single probability forecast. 

\subsection{Scoring}

The accuracy of submitted forecasts were measured using Brier scores~\cite{brier_g_w_verification_1950}, the squared distance of the forecast from the result, coded as 1 if the event/quantity was realized, and 0 otherwise. \rev{We use Brier scores to measure accuracy because they are specifically designed to assess the accuracy of probabilistic information (unlike other metrics, like F1). As a variation of squared-error, Brier scores penalize more egregious errors more severely. In addition to scoring accuracy, a Brier score is also a proper scoring rule meaning it incentivizes responding honestly. Practically, the HFC test and evaluation team chose Brier scores as their primary accuracy metric. Using the same metric allowed us to efficiently track our performance compared to control and the other HFC competitors. Briers scores can be interpreted similarly as mean squared error. A minimal baseline for accuracy is to show improvement over an uninformed judge, who assigns equal probabilities to all C bins (prob = 1/C), earns a Brier score of (C-1)/C. Brier scores are also commonly described as improvement over a known comparator. Below we compare to control using Cohen's d, a standardized mean difference, and in some instances display the percent improvement.

We used formulations of the Brier score based on the number of response options and ordinality of the IFP~\cite{merkle_weighted_2018}. This Brier score variant ranged from 0 (perfect accuracy) to 2 (worst possible score).} The accuracy of each forecasting slot for a given IFP was characterized by the Mean Daily Brier score (MDB), e.g., the Brier score averaged over the active days of that IFP. Usually, the SAGE system submitted daily forecasts for each open IFP. If for whatever reason a forecast was not submitted on any given day (e.g., system outage), the last submitted forecast was carried forward. If a slot did not submit any forecasts at all for a given IFP, a uniform prior was used to calculate the score. 

A similar approach was used to score individual forecasters. A forecast for a given user was carried forward until that user chose to revise the forecast. If a forecaster did not place an estimate on the first day of a question, we imputed the median score across all forecasters in a condition for each day an IFP was open prior to the first forecast. A user's score across IFPs was the mean of MDBs or MMDB. To adjust for the difficulty of individual IFPs and aid in interpreting comparisons across conditions, we standardized Brier scores to have a mean of zero and standard deviation of 1 for each IFP-day.

\section{Results} \label{results}
\subsection{Aggregate Performance}
First, we report our main result that compares the  aggregate performance of the SAGE system with the non-hybrid control. As we mentioned above, each method was allocated 40 official and 60 experimental slots for submitting aggregated forecasts. Table \ref{tab:main} summarizes our results for both official and experimental slots across 398 IFPs. The best official SAGE method led to an improvement in mean accuracy of a Cohen's d of 0.126 over control. 

\begin{table}[!h]
    \centering
    \begin{tabular}{|c|c|c|}
    \hline
    \textbf{Condition} & \textbf{Official} & \textbf{Experimental} \\
    \hline
         Best Performing Control  & 0.3398 & 0.3325 \\
         \hline
        Best Performing SAGE & 0.3065 & 0.3052 \\
        \hline
    \end{tabular}
    \caption{The Brier scores of the best performing official and experimental methods for both SAGE and the control.}
    \label{tab:main}
\end{table}

SAGE's best-perfoming aggregation slot had the following properties. First, it used both Control and SAGE human forecaster data as inputs. Second, it applied a time-varying weighted mean human aggregation algorithm, which made forecaster weights more unequal over time; aggregate forecasters were de-extremized at the start of the season and extremization parameter value increased over time, resulting in light extremization by the end of the season. Third, human and machine-model forecasts were combined using a rule that produced a weight of a machine model forecast as equivalent to eight average-skill human forecasters on time-series questions, and four average-skill human forecasters on other questions that used less sophisticated models.  

SAGE outperformed the control condition both for the official and experimental slots. We ran backcasting analyses to estimate the impact that different aspects of our system, including SAGE forecasters, human aggregation, and machine forecasts, contributed to SAGE outperforming the best Control method, a Brier score difference of 0.0333. Results showed that applying the SAGE human-aggregation algorithm to control human forecasts would have resulted in a Brier score advantage of 0.01, approximately 31\% of the full difference. In retrospect, the distinguishing feature of the best-performing human aggregation model was that it extremized aggregate forecasts less, especially at the start of the season. Applying this human-aggregation algorithm to the combination of control and SAGE human forecasts resulted in a Brier score advantage for SAGE of 0.032, approximately 97\% of the full difference. The further addition of machine models at the aggregation stage rounded up the full 100\% advantage. For more details on the accuracy benefits of machine models at the user interface vs. aggregation stage (see Section \ref{scalable}).

\subsection{Individual Performance Across Conditions}
We analyzed user performance in the various conditions across the 398 resolved questions. Table \ref{tab:condpops} lists the volume of users and forecasts in each condition.
\begin{table}[!h]
    \centering
    \begin{tabular}{|c|c|c|c|}
    \hline
    \textbf{Condition} & \textbf{Users} & \textbf{Forecasts} & \textbf{Forecasts/User}\\
    \hline
         B & 190 & 25,163 & 132.4 \\
         \hline
        C & 158 & 20,782 & 131.5 \\
        \hline
        D & 199 & 27,348 & 137.4 \\
        \hline
        Total (SAGE) & 547 & 73,293 & 134.0 \\
        \hline
        A (Control) & 538 & 79,611 & 148.0 \\
        \hline

    \end{tabular}
    \caption{Number of unique forecasters and generated forecasts in each experimental condition.}
    \label{tab:condpops}
\end{table}

Since some questions were relatively easy and highly predictable and others were more difficult, we expected them to yield (possibly, very) different Brier scores. Thus, whenever comparing, or aggregating, Brier scores across multiple items, it was important to adjust for inherent imbalance in difficulty. Our approach to this problem was to standardize the Brier scores for every question to have a mean of 0 and a SD of 1, across all the responses in all conditions, before combining them. Thus, we report results in terms of mean, median, and 25th percentile standardized Brier scores. The lower (and more negative) a score is, the more accurate it is. The results of all conditions are presented in Table~\ref{tab:standardizedperf}.

\begin{table}[!h]
    \centering
    \begin{tabular}{|c|c|c|c|c|}
    \hline
    \textbf{Condition} & \textbf{Mean} & \textbf{Std. Dev.} & \textbf{25th Percentile} & \textbf{Median} \\
    \hline
    A (Control) & \textbf{-0.021} & 0.958 & -0.486 & \textbf{-0.259} \\
    \hline
    B (Data) & 0.051 & 0.987 & -0.472 & -0.156 \\
    \hline
    C (Models) & 0.039 & 1.143 & \textbf{-0.537} & -0.198 \\
    \hline
    D (Interactive) & 0.002 & 0.989 & -0.508 & -0.196 \\
    \hline

    \end{tabular}
    \caption{Comparison of standardized (at IFP level) Brier scores across conditions including mean, median, and 25\% percentile for each. Bolded values represent the lowest (most accurate) score within each column.}
    \label{tab:standardizedperf}
\end{table}

Our results indicated that users in conditions C and D outperform those in condition B, but only the most skilled forecasters outperformed the control (no data) condition. We confirmed the hypothesis that having access to model predictions indeed helps \textit{skilled}, but not \textit{average}, forecasters. The greatest improvement came when data charts were available, and skilled forecasters viewed model predictions, z-Brier = -0.618 vs. -0.545 for control. Note that the availability of more models and interactive features, provided in condition D, did not necessarily help with performance. Indeed, while condition D had a better mean score across all the questions, users in this condition did not perform well on questions where data charts were available. This suggests that the availability of multiple models and/or interactive features do not help the users to generate more accurate forecasts. In fact, users used the various options available to them very rarely.

\rev{Here we present overall performance of our system and experimental conditions. For an analysis of individual conditions, see ~\cite{himmelstein_forecasting_2021}. We only find consistent differences by major IFP format, but not by traits like topic area, region, or question duration. Our model was most accurate for binary questions, zBrier = -0.107 (SD = 1.10), and least accurate for non-ordinal IFPs, zBrier = 0.180 (SD = 0.97). As discussed in Section \ref{agg methods}, we did include certain IFP and linguistic traits when testing more complex aggregation methods, but in most instances, they underperformed our interpretable aggregation methods except in the anchor attention model~\cite{huang_anchor_2022}.}

\subsection{Model-Based Forecasts}
\label{sec:results-models}

We also analyzed the performance of the machine models outlined in Section~\ref{sec:methods-models}. Overall, simple ensemble models worked well. The best performing model (``PHE2'') was an ensemble that averaged the forecasts from Auto ARIMA and an exponential smoothing state-space model (ETS)~\cite{hyndman2018forecasting}. It slightly outperformed the M4-Meta model~\cite{montero2020fforma} that ranked 2nd highest in the M4 time series forecasting competition, and clearly outperformed more complex methods like a recurrent neural network and custom-coded regularized auto-regressive model. Even the Auto ARIMA model itself did reasonably well throughout. From a practical standpoint, the simpler ensembles were computationally less expensive, had fewer software dependencies, and were less likely to break.

\begin{figure}[!h]
    \centering
    \includegraphics[width=0.6\textwidth]{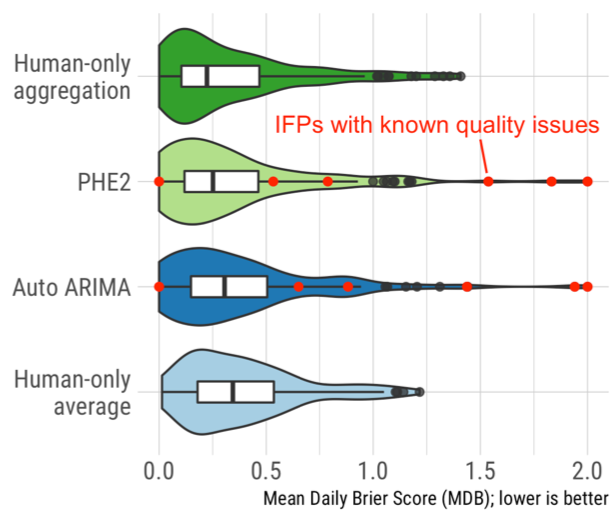}
    \caption{Relative performance of two model-based forecasts compared to average human performance and the best human forecast-only aggregation model. Auto ARIMA was a mainstay model throughout; PHE2 emerged later as a top performer. This figure includes performance on 153 IFPs for which all models had forecasts. Red points mark IFPs with known quality issues that were retained for the sake of coverage. }
    \label{fig:mm_fig}
\end{figure}

Model forecasts relative to the human forecasts were overall near or at parity with human forecasts. Figure~\ref{fig:mm_fig} shows the distributions of mean daily Brier scores for the Auto ARIMA and PHE2 models, as well as a simple average of human forecasts and the best-performing aggregation model of human forecasters from condition B, which were not exposed to machine model predictions. Both models outperformed the average human forecast but lagged slightly behind the best aggregation model of human-only forecasts. In part, this is because they had a small number of very bad forecasts. Some of these were caused by IFPs with known data quality issues, which tended to lead to extreme forecasts with either very low or very high Brier scores. Some of these---the more easily identifiable ones---are marked with the red points. Lastly, despite similar average performance, the model and human forecasts did well or poorly on different questions. For example, the inter-question correlation of performance for the PHE2 and aggregation benchmark models was only 0.3. 

After enough IFP results were observed, we developed a meta-model for the relative performance of the time-series model and human forecasts. Overall, there were no clear bivariate or multivariate relationships between a large variety of question and data features and the relative model to human forecast performance, which for example could have led us to identify a subset of IFPs in which one consistently outperformed the other. However, towards the end of our experiment, enough performance data had accumulated so that a random forest model trying to identify forecasts that were clearly worse than a uniform forecast, or forecasts that can beat the aggregation benchmark, achieved slightly informative accuracy levels, with out-of-sample AUC-ROC values of 0.69 and 0.60 respectively. This may have been sufficient to implement a filter for likely bad forecasts, if the experiment continued.

\subsection{Using Machine Models for Scalable Forecasting} \label{scalable}

We analyzed aggregate performance on the IFPs for which our strongest machine model was available, a discrete-cosine transform (DCT) ensemble. We found incorporating machine models during aggregation led to improvements at several stages which accounted for our team's overall advantage. Although these accuracy gains were consistent throughout the competition, effects at individual stages were modest and none were statistically significant on their own. The results showed that providing forecasters access to model projections led to modest improvements in aggregate accuracy. Forecasters who could view model forecasts before making their estimates produced aggregate forecasts with 6\% better Brier scores, compared to aggregations of forecasters with no access to model projections. Injecting model estimates at the aggregation stage also led to small improvements in accuracy (i.e., reductions in Brier score) of 2\%-3\% points.
\ignore{
\begin{figure}[!h]
    \centering
    \includegraphics[width=0.8\textwidth]{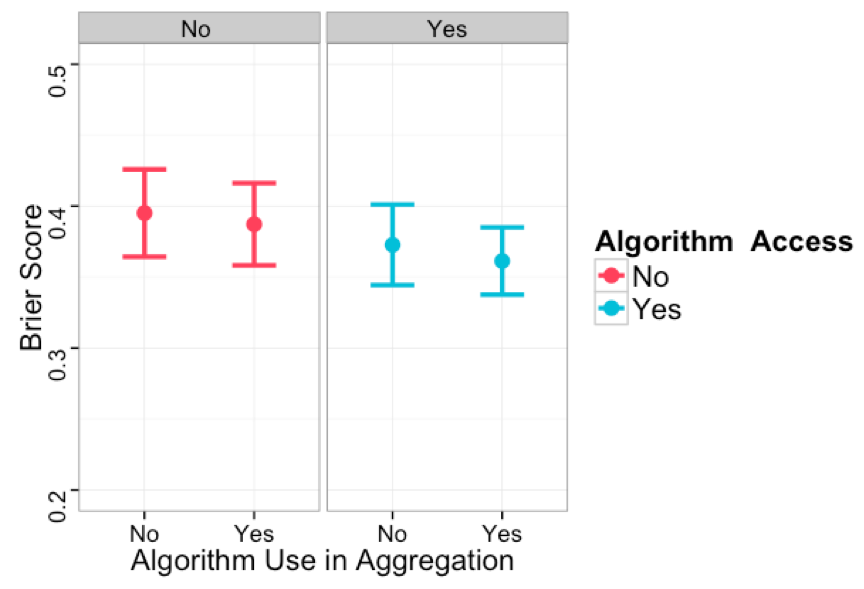}
    \caption{Average aggregate performance (Brier score) as a function of forecaster access to model projections through the user interface (Yes, No), and inclusion of machine model forecasts at the aggregation stage. Higher Brier scores correspond to worse aggregate accuracy. Whiskers denote 1 standard error of the mean.}
    \label{fig:agg_avgaggperf}
\end{figure}
}
\begin{figure}[!h]
    \centering
    \includegraphics[width=0.8\textwidth]{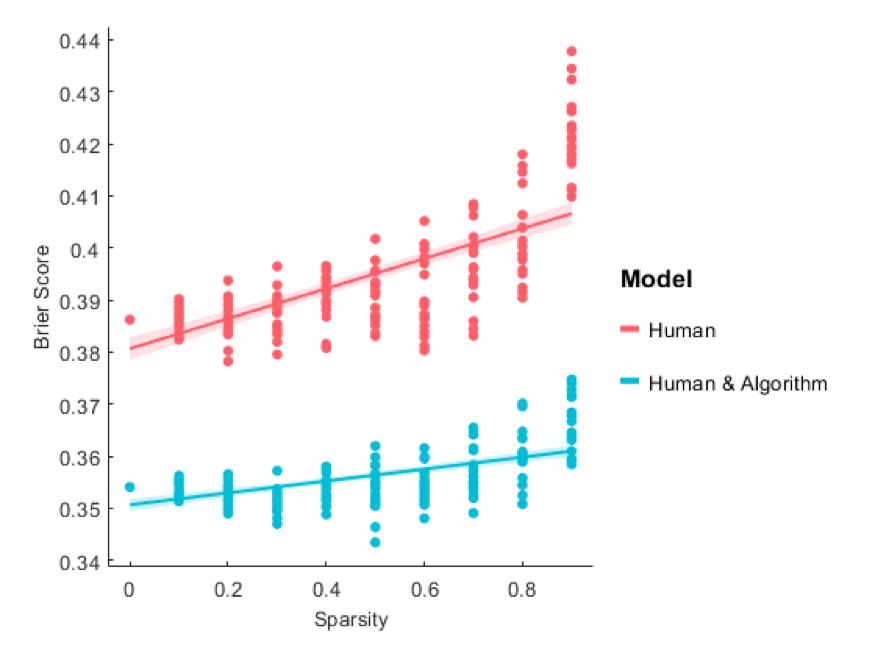}
    \caption{Average aggregate performance (Brier score) as a function of the proportion of human forecasts removed from the forecasting pool (Sparsity). Higher Brier scores correspond to worse aggregate accuracy. Each point corresponds to aggregate performance for a random subset of censored forecasts. The line plots the linear regression of these points and the shaded region is the 95\% confidence interval based on N=20 simulations.}
    \label{fig:agg_proportion}
\end{figure}

The benefits of including model forecasts in the aggregate became more salient when considering the issue of scale. Scaling up the number of questions in a human-only pool means fewer human forecasts for each question, which is expected to degrade aggregate performance. To simulate this effect, we made human forecasts sparser by deleting a random subset of users from the aggregate. The results demonstrated that including model forecasts insulated the aggregate forecast against the negative effects of sparse human judgments that would occur when scaling to large numbers of questions (see Figure~\ref{fig:agg_proportion}).

\subsection{Impact of Training}
Participants were randomly assigned to either a brief (about 30 minute) training or a control condition in which they read popular articles about forecasting, but without tips for boosting accuracy. Training occurred once, during their second active week, and was accessible for review on the platform menu. We assessed accuracy and activity to see if trained forecasters worked harder than untrained users. We first compared forecast accuracy before and after training exposure to ensure trained forecasters were not randomly better from the start. We found trained forecasters outperformed control forecasters post-exposure (d = 0.56). We further assessed whether accuracy could be improved via the delivery method of the training material. Forecasters who saw animated material significantly outperformed forecasters who saw static material in average accuracy (t(410) = 3.55, p $<$ .001), generated slightly more forecasts per IFP than the static group, (d = 0.20, t(342) = 1.98, p = .049), and attempted approximately 5\% more IFPs than static-trained counterparts, a significant difference (d = 0.23, t(342) = 1.98, p = .018). More details can be found in ~\cite{Joseph2019}.  

\subsection{IFP Recommendation}
\begin{figure}[!h]
    \centering
    \includegraphics[width=0.6\textwidth]{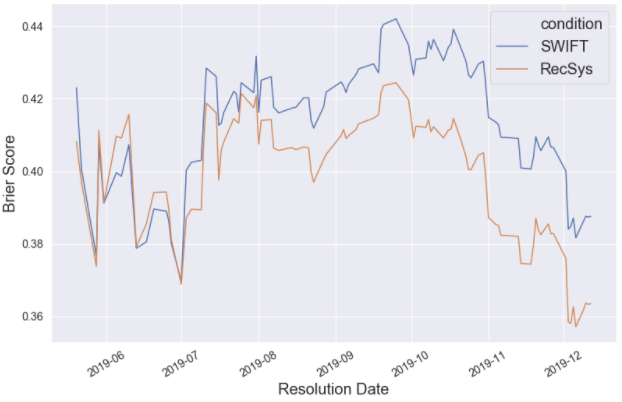}
    \caption{The IFP Recommender System learns over time the skills and preferences of each forecaster. After 2 months into the experiment, the forecasters who received the recommendations start to consistently outperform the forecasters in the control condition (SWIFT), with a relative improvement that stabilizes around 7\%.}
    \label{fig:ifprecsys}
\end{figure}

We measured the performance benefit of users assigned to questions ordered using the IFP Recommender System, described in Section ~\ref{sec:matching}, vs. the global ranking based on resolution date and popularity (SWIFT ordering). As shown in Figure~\ref{fig:ifprecsys}, we obtained a statistically significant improvement, up to a 7\% relative decrease in Brier score by the end of the experimental phase.

\begin{table}[!h]
    \centering
    \begin{tabular}{|c|c|c|}
    \hline
    \textbf{Method} & \textbf{Brier Score} & \textbf{Budget (\% of forecasts)} \\
    \hline
         All forecasts & 0.397 & 100 \\
\hline
Random forecasts & 0.402 & 62 \\
\hline
GreedyIFP & 0.376 & 61.4 \\
\hline
GreedyIFP++ & 0.375 & 37.8 \\
\hline
    \end{tabular}
    \caption{Brier score of different IFP recommender systems. All results are averaged over 10 runs.}
    \label{tab:recsysbrier}
\end{table}

We further assessed the IFP Recommender System in terms of effective resource allocation by simulating different allocation strategies. First, since we could not foresee who the best performers would be before the phase ends, we implemented a greedy approach to improve our cohort of users by periodically excluding a certain percentage of worst performers every time a batch of IFP closed (i.e. got resolved) (termed \textbf{GreedyIFP}). Second, we additionally capped the number of forecasts on popular IFPs to reallocate forecasts of diminishing return after consensus was reached (termed \textbf{GreedyIFP++}). As shown in Table~\ref{tab:recsysbrier}, both greedy strategies obtained a small improvement in the global Brier score, while reducing the forecasting budget respectively by 38\% and 62\%. In contrast, we showed naively decreasing the number of forecasts (``Random forecasts'') negatively affected the global Brier Score.

\section{Discussion}
In the above, we show how a hybrid forecasting system can outperform established crowd-sourced forecasting systems. The SAGE hybrid system consistently outperforms the human-only control condition. Our hybrid system improves the accuracy in the aggregate, although improvements were modest. Beyond the proven methods for aggregating human forecasts~\cite{atanasov_distilling_2017}, we find the inclusion of machine models in the user interface and the aggregation algorithms was key to improving accuracy. Individually, access to model predictions only improves the accuracy of highly skilled forecasters. While this is evidence of their value, it provides further evidence that forecasters must have enough expertise to know when and how to use this information~\cite{abeliuk_quantifying_2020}. Predicting the future is difficult, especially for deeply uncertain, impactful geopolitical events. Humans and machines are both limited by the irreducible uncertainty of the setting. Combining human and machine predictions leads to gains in accuracy by helping protect against some of the most egregious errors, especially when the two sources disagree.  

In addition to better forecasting accuracy, another critical advantage of SAGE over crowdsourcing-based systems is its scalability. We find evidence that the SAGE hybrid system helps answer more questions with the same number of human users without losing accuracy, although the scope of these improvements remains an open hypothesis. Adaptive question-user assignment increases the ability to scale by limiting the number of users who can access each IFP once consensus is achieved. Our recommender system succeeds based on three primary features. First, it provides a unique question ranking for each user based on the IFPs they previously chose to answer. Second, it excludes users who tend to forecast early and perform poorly. Third, it identifies an optimized number of users per IFP and capped each IFP that exceeded that maximum, thus shifting forecasts from popular to unpopular IFPs and increasing the utility of post-consensus forecasts. 

Our results are subject to a number limitations, most of which are a function of participating in a forecasting tournament managed by a third-party \textit{test and evaluation} team. Notably, our machine models needed to be robust and flexible since new datasets and question formats were regularly published unannounced. Our system was successfully able to ingest large amounts of, often unformatted, data in the window between question publication and user recruitment - often just a few hours. Thus, our results might not generalize to \rev{more stable} situations with highly developed models tuned to a specific environment or dataset.

We highlight two keys lessons about the model contribution to the ``hybrid'' system. First, aim for depth, not breadth. Our initial strategy, in the spirit of the hybrid part of the competition, was to try to cover as many IFPs as possible. This led to several decisions to use marginal or non-canonical data. The quality issues tended to lead to extreme forecasts that sometimes were really good (achieved a low Brier score), but more often were really bad and thus reduced average quality. A better strategy would have been to focus on a smaller subset of questions where good performance can be achieved, and spend more time on quality control rather than coverage. Better average quality also simplifies downstream use of forecasts in aggregation.  

Second, data is king (or more familiarly, ``garbage in, garbage out''). The main cause of poor model forecasts was data quality issues. Some sources, like the OECD~\cite{OECD_nodate} and OPEC~\cite{OPEC_nodate}, alter historic data values when updating. There were also many questions that required data transformations or had marginal data ill-suited for time-series. For example, count series based on transformed ACLED event data~\cite{ACLED_nodate} were plagued by inaccuracies due to idiosyncratic technical issues in the data platform back-end. Unlike errors at the modeling stage, these kinds of data problems usually were hard to identify without labor-intensive manual reconstruction of a series from its source. Selecting good-enough time series models, which had been the focus of our efforts in the beginning, in the end turned to be easier than these two issues. 

\rev{Similarly, the success of our simpler, interpretable aggregation methods is most likely due to model training. The greater complexity of the aggregation method, the more training data it required. Complex aggregation methods suffered from training during burn-in due to limitations in the number of IFPs that resolved earlier in the season and variability between sources in format, frequency, and availability of data. Such methods suffered more inefficiency due to retraining mid-season due to the requirement that our system be able to handle newly introduced data sources on the fly. In complex settings, simpler, traditional statistical methods often outperform novel, complex methods~\cite{makridakis_statistical_2018}. Research on forecast combinations supports the success of simple conservative methods~\cite{armstrong_golden_2015}.}    

The recruitment of human forecasters for such long-term engagement is also innately challenging. Recruitment was managed by a third party according to HFC rules. This system prioritized retention over accuracy incentives, and HFC rules limited our ability to add performance-based incentives beyond those offered by the recruitment team. Changing the retention-accuracy incentive balance is likely to alter the quality of human performance.

In conclusion, gains from hybridizing are consistent, but modest in this setting. \rev{The SAGE system's success relies on both computer-in-the-loop hybridization} including the information \rev{(historical data and model predictions)} shown to users and mandated \rev{narrative graphical} training, \rev{as well as human-in-the-loop hybridization including human inputs into the aggregation algorithms} and strategic user-IFP assignment, to name a few. It is important to engineer such a complex system to optimize the interactions between each component, since each improves accuracy slightly. \rev{The optimal system must balance several tradeoffs, like using models that are no more complex than the tuning parameters that can be confidently estimated, and anchoring users on objective, data-driven benchmarks while eliciting the diversity required for crowd wisdom.} The real advantage is not boundless improvements in accuracy. Instead it is the ability to tackle a greater burden without needing to increase human resources. When human-question balance is sparse, it is important to view users as a labor pool and use adaptive question assignment to maximize human coverage. 

\section*{Acknowledgments}
The authors would like to thank Seth Goldstein, Peter Haglich, Rob Hartman, Daniel Horn, and Steven Rieber for their helpful feedback during the HFC program. This research is based upon work supported in part by the Office of the Director of National Intelligence (ODNI), Intelligence Advanced Research Projects Activity (IARPA), via 2017-17071900005. The views and conclusions contained herein are those of the authors and should not be interpreted as necessarily representing the official policies, either expressed or implied, of ODNI, IARPA, or the U.S. Government. The U.S. Government is authorized to reproduce and distribute reprints for governmental purposes notwithstanding any copyright annotation therein.

\bibliographystyle{plain}

\end{document}